\documentclass[12pt]{article}

\usepackage{amssymb, amsmath}

\begin{document}

\title{\begin{flushright}
\vspace*{-3.3cm}
{\small{UAB-FT-676,  ULB-TH/09-42}}
\end{flushright}
\vspace{+0.9cm}
Dark matter stability and unification\\ without supersymmetry}
\author{Michele Frigerio$^{a,b}$, Thomas Hambye$^c$}
\date{}
\maketitle
\begin{center}
{\sl $^a$ Institut de F\'isica d'Altes Energies, Universitat Aut\`onoma de Barcelona, \\
E-08193 Bellaterra, SPAIN}\\
{\sl $^b$ Institut de Physique Th\'eorique, CEA-Saclay,\\ F-91191 Gif-sur-Yvette Cedex, FRANCE}\\
\indent {\sl $^c$ Service de Physique Th\'eorique, Universit\'e Libre de Bruxelles,\\ 1050 Bruxelles,
BELGIUM}\\
\end{center}

\abstract{\sf
In the  absence of low energy supersymmetry, we show that
(a) the dark matter particle alone at the TeV scale can improve gauge coupling unification,
raising the unification scale up to the lower bound imposed by proton decay,
and (b) the dark matter stability  can automatically follow from the grand unification symmetry.
Within reasonably simple unified models, a unique candidate satisfying these two properties
is singled out: a fermion isotriplet with zero hypercharge, member of a $45$ (or larger) representation of $SO(10)$.
We discuss the phenomenological signatures of this TeV scale fermion, which can be tested
in direct and indirect future dark matter searches.
The proton decay rate into $e^+\pi^0$ is predicted close to the present bound.
}

\section{Introduction}

Soon the Large Hadron Collider (LHC) will explore the origin of the electroweak symmetry breaking and possible new physics at the TeV scale.
Implications for the theory at
higher energy scales will certainly be profound.
While one Higgs doublet suffices to account for the spontaneous symmetry breaking
and for the present electroweak data, the standard model (SM) alone does not answer several questions. In particular it does not provide 
a dark matter (DM) candidate,
nor the extra ``light" states required to raise the weak-electromagnetic unification scale, in order to sufficiently suppress the proton decay.

While low energy supersymmetry is an attractive completion of the SM that 
could address these issues and be discovered at the LHC,
so far supersymmetry has not been observed at scales as low as expected to fully cure the hierarchy problem, and it requires
additional theoretical assumptions to be viable phenomenologically.
Therefore, even if not solving the hierarchy problem, non-supersymmetric completions of the SM at the TeV scale
should be seriously contemplated to address more phenomenological issues.

One piece of new physics which is highly motivated at the TeV scale
is the DM particle $\--$ it is needed if the DM relic density follows from the thermal freeze-out of its annihilation (the WIMP mechanism).
A legitimate question one could ask is whether
such a DM candidate could be precisely the state missing for gauge unification.
Furthermore,
in most models the origin of the DM stability against decays into SM particles is not clear.
An ad hoc discrete symmetry, e.g.~a $Z_2$ parity, is assumed.
Another sensible question one could ask  is whether this global symmetry could derive from a more motivated 
gauge grand unification theory (GUT).
This, in turn, would have the advantage to prevent quantum gravity effects from breaking this symmetry \cite{KW}.
In the following we will show that a positive answer to these two questions can be provided within a simple framework without invoking supersymmetry.

\section{Stable particles in $SO(10)$ unification}

We begin by exploring the possibility to obtain a stable DM candidate without introducing,
on top of the unified group, any {\it ad hoc} discrete or continuous symmetry. 
The simplest way to preserve a $Z_2$ subgroup of the unified group $G$ is to break spontaneously a $U(1)$ subgroup of $G$ with fields
carrying an even $U(1)$ charge. For this purpose $G$ should have rank 5 or larger,
in order to contain an extra $U(1)$ factor besides the SM group. 
The most straightforward possibility arises if we break the $U(1)_{B-L}$ subgroup of $SO(10)$
by the vacuum expectation values (VEVs) of fields with even $B-L$.
This preserves the discrete symmetry $P_M = (-1)^{3(B-L)}$, known as {\it matter parity}.
In the context of supersymmetric models, $P_M$ can be identified with $R$-parity and 
the possibility to automatically obtain it from a left-right symmetric (unified) gauge theory was recognized in Ref.~\cite{rparity}.
More recently, the $SO(10)$ origin of $P_M$ was employed in non-supersymmetric DM models
\cite{KKR}.

All the components of the same $SO(10)$ multiplet carry the same matter parity.
The representations of dimension $16$ and $144$ have $P_M=-1$ (odd), while all other representations
with dimension $\le 210$ have $P_M=1$ (even).
If the SM fermions are part of $16_i$ multiplets for $i=1,2,3$, and the SM Higgs is part of a $10$ multiplet
(or any other even multiplet), then any new {\it odd scalar} or {\it even fermion} cannot decay into SM particles only,
because of
unbroken $P_M$. The lightest of these new particles  is then automatically stable.

The simplest candidates for {\it scalar} DM are thus the neutral and colourless components of a $16$ scalar multiplet:
the SM singlet $S^{16}$ and the neutral component of the isodoublet $D^{16}$. 
The phenomenology of the singlet scalar DM \cite{McDonald1}  as well as of the inert doublet DM \cite{Deshpande:1977rw}
has been studied in detail.
The $SO(10)$ origin of their stability has been pointed out recently \cite{KKR}.
Some model-building issues for these scalar DM candidates are discussed in Ref.~\cite{KKR2}.
The scalar DM candidates contained in odd $SO(10)$ representations larger than $16$
are displayed in Table \ref{table}.

The candidates for {\it fermion} DM belonging to the smallest even $SO(10)$ representations are 
the isodoublets contained in a $10$ multiplet, the $Y=0$ isotriplet contained in $45$ or in $54$, the $Y=\pm1$ isotriplets contained in $54$.
The multiplets $45$ and $54$ also contain SM singlets, which however do not interact with the SM at the renormalizable level and
consequently require extra relatively light states, in order to couple these singlets to the SM and thus lead to viable
DM candidates.
The fermion DM candidates contained in $SO(10)$ representations larger than $54$
are also displayed in Table \ref{table}.

The case of a fermion DM candidate presents a few advantages: 
(i) since a Weyl fermion contributes four times more than a real scalar to the gauge coupling evolution,
a unique and small DM multiplet can significantly improve gauge unification, as shown in section \ref{unif};
(ii) there is no need of odd scalar multiplets in the model
$\--$ in their absence the matter parity $P_M$ is automatically conserved, otherwise
one needs to assume that they do not acquire a VEV;
(iii) the lightness of fermion DM in the effective theory below the unification scale $M_{GUT}$
is natural in the 't Hooft sense, 
since a global $U(1)$ symmetry appears in the limit of vanishing DM mass (on the contrary, a light  scalar DM
in the absence of supersymmetry would require an extra fine-tuning with respect to $M_{GUT}$).
A detailed discussion of $SO(10)$ model-building is postponed to section \ref{trip}.

\begin{table}[t]
\begin{center}
\small{
\begin{tabular}{|l||c|r||c|r|}
\hline
& \multicolumn{2}{|c||}{fermions} & \multicolumn{2}{|c|}{scalars} \\ \hline
\begin{tabular}{c} $SU(2)_L\times U(1)_Y$\\ representation \end{tabular} &   
\begin{tabular}{c} even $SO(10)$ \\ representations \end{tabular} & $b_1^{DM}-b_2^{DM}$  & 
\begin{tabular}{c} odd $SO(10)$ \\ representations \end{tabular} & $b_1^{DM}-b_2^{DM}$ \\ \hline
\hline
\qquad\quad$1_0$  & 45, 54, 126, 210 & 0 & 16, 144 & 0\\
\hline
\qquad\quad$2_{\pm1/2}$  &  10, 120, 126, 210, $210'$ & $-4/15$ &  16, 144 &  $-1/15$ \\
\hline
\qquad\quad$3_0$ & 45, 54, 210 &   $-20/15$ & 144 & $-5/15$ \\
\hline
\qquad\quad$3_{\pm1}$  & 54, 126 &  $-4/15$ & 144 &  $-1/15$\\
\hline 
\qquad\quad$4_{\pm1/2}$  &  $210'$ &  $-88/15$ & 560 & $-22/15$ \\
\hline
\qquad\quad$4_{\pm3/2}$  & $210'$ & $+8/15$ & 720 & $+2/15$ \\
\hline 
\qquad\quad$5_0$  & 660 & $-100/15$   & 2640 & $-25/15$ \\
\hline 
\qquad\quad\dots &  \dots & \dots & \dots & \dots \\
\hline 
\end{tabular}}
\end{center}
\caption{\it{\small{The electroweak multiplets containing a neutral component are listed in the first column. 
In the second (fourth) column, we display the even (odd) $SO(10)$ representations
that contain one or more multiplets with the given electroweak charges and no colour. 
We list all representations with dimension $\le 210$ or, when there are only larger representations, the smallest one.
In the third (fifth) column, we give the contribution to the gauge coupling $\beta$-function coefficients, computed
for the minimal number of degrees of freedom: when $Y=0$ one Weyl fermion (one real scalar), 
when $Y\ne 0$ two Weyl fermions (one complex scalar).}}
\label{table}}
\end{table}

\section{Gauge unification in the standard model\\ plus dark matter \label{unif}}

It is worth to ask whether one of the naturally stable DM candidate above, with mass at the TeV scale as required by the WIMP thermal freeze-out mechanism, could also account for gauge coupling unification. 
In the SM with one complex Higgs doublet, the hypercharge and weak  gauge couplings $\alpha_1$ and $\alpha_2$ 
meet each other at $M_{GUT}^{SM}\simeq 10^{13}$ GeV.
This cannot be the true grand unification scale: (i) the inferred value of the strong coupling, $\alpha_3^{SM}\simeq 0.07$,
is by far smaller than the experimental value, $\alpha_3^{exp} = 0.118(2)$;
(ii) the proton decay is far too fast, since the $SU(5)$ gauge bosons should have a mass $M_V\gtrsim 4 \cdot 10^{15}$ GeV
to comply with the present bound on the proton lifetime, $\tau(p\rightarrow \pi^0e^+) > 8.2\cdot 10^{33}$ years at $90\%$ \cite{SK}.

The scale $M_{GUT}$ where $\alpha_1=\alpha_2$ can be raised with respect to the SM value
by extra multiplets with mass at intermediate scales.
At one loop, one finds
\begin{equation}
\log \frac{M_{GUT}}{M^{SM}_{GUT}} = - \sum_a \frac{b_1^{(a)}-b_2^{(a)}}{b_1-b_2} \log \frac{M^{SM}_{GUT}}{m_a} ~, 
\end{equation}
where the sum runs over non-standard multiplets with mass $m_a$, $b_i^{(a)}$ is the corresponding coefficient in the
$\beta$-function for $\alpha_i$ and we define $b_i = b_i^{SM} + \sum_a b_i^{(a)}$, with $b_1^{SM}=41/10$, $b_2^{SM}= -19/6$ 
and $b_3^{SM} = -7$.
For orientation, in order to obtain $10^{19}$ GeV $> M_{GUT} > 10^{15}$ GeV with a multiplet of mass $m_{DM}= m_Z=91.2$ GeV, one needs
\begin{equation}
-2.6 < b_1^{DM}-b_2^{DM} < -1.1 ~.
\label{window}\end{equation} 
The smallest electroweak multiplets with a neutral component are listed in Table \ref{table}, 
together with the corresponding value of $b_1^{DM}-b_2^{DM}$.
The value in the Table corresponds to  the minimal number of degrees of freedom, that is to say, one real scalar
or one Weyl fermion in the case of multiplets with zero hypercharge,  one complex scalar or two Weyl fermions
in the case of multiplets with $Y\ne 0$.

If one sticks to such a minimal field content, 
we find that the isotriplet with $Y=0$, denoted as $3_0$, is the only fermionic DM candidate that fits into the window given in 
Eq.~(\ref{window}). There are also two scalar DM candidates in this window, namely $4_{\pm1/2}$ and $5_0$,
but they are contained in huge $SO(10)$ representations only, the smallest with odd matter parity being
the $560$ and the $2640$, respectively. This shortcoming applies also to all larger electroweak
multiplets, not displayed in Table \ref{table}. In the following we will therefore consider the simplest possibility, a $Y=0$ isotriplet 
$T\equiv (T^+,T^0,T^-)$ and, taking into account the discussion of section 2, 
we will assume it to belong to the $45$ (or $54$, $210$, \dots) $SO(10)$ representation. 

We just mention the alternative option to admit several copies of the same multiplet or the coexistence of different DM candidates.
For example, the set of fermions $(3_0,2_{\pm 1/2})$ would mimic the case of wino plus higgsinos in supersymmetric models. 
Another variation could be to include at TeV scale one fermion $3_0$ as well as one scalar $3_0$. 
This spectrum appears in a class of $SU(5)$ models with low scale seesaw, which contain one fermion and one scalar $24$ multiplet 
\cite{BS,DF}; in these models no stable particle exists that can play the role of DM \cite{BNS}, unless an extra symmetry is added
by hand \cite{FIR}.
The unification predictions as well as the DM phenomenology are much less constrained in these scenarios
with multiple DM candidates, and we will not discuss them any further.

Besides modifying the point where $\alpha_1$ and $\alpha_2$ meet, 
the intermediate scale multiplets also affect the prediction for $\alpha_3$ as follows:
\begin{equation}
\frac{1}{\alpha_3(m_Z)} = \frac{1}{\alpha_{GUT}} + \frac{b_3^{SM}}{2\pi}\log\frac{M_{GUT}}{m_Z} +
\sum_a \frac{b_3^{(a)}}{2\pi} \log\frac{M_{GUT}}{m_a} ~.
\label{as}
\end{equation}
The increase of $M_{GUT}$ with respect to the SM, as required to suppress proton decay,
and the corresponding increase of $\alpha_{GUT}$
go in the right direction, since they raise the SM prediction for $\alpha_3(m_Z)$.
Unfortunately for $b_3^{DM}=0$, that is the case for a colourless DM particle, one actually overshoots the experimental value:
$M_{GUT}>10^{15}$ GeV implies $\alpha_3(m_Z) \gtrsim 0.17$, independently from the value of $m_{DM}$.
The experimental value of $\alpha_3$ indicates that coloured multiplets with  $b_3^{(a)}>0$ are also present below the GUT scale.

The nature of the coloured multiplets below the GUT scale depends of course on the choice of the DM candidate. 
When $T$ is added to the SM, the value of $M_{GUT}$ turns out to be close to the lower bound
imposed by the proton lifetime (a detailed discussion is given in section \ref{pheno}).
Therefore,
among the coloured components of small $SO(10)$ representations (dimension $\le 54$), one should better select those with
$b_1^{(a)}-b_2^{(a)}\le 0$, otherwise they would further lower $M_{GUT}$.
There are two
such multiplets, with SM quantum numbers $(3,2,1/6)+(\bar{3},2,-1/6)$ and $(8,1,0)$. 
Their contribution to the $\beta$-function coefficients read 
$(b_1^{(a)},b_2^{(a)},b_3^{(a)})=(2/15,2,4/3)$ and $(0,0,2)$ respectively, in the case they are fermions.
Both are an economical choice, since they belong to the same $45$ (or $54$) $SO(10)$ multiplet as $T$.
However it turns out that these colour triplets do not lower the prediction for $\alpha_3(m_Z)$, because
the effect of the third term in Eq.~(\ref{as}) is more than compensated by the decrease of the first two terms.

We are thus left with the colour octet $O$, which does not modify $\alpha_{GUT}$ nor $M_{GUT}$
and can account for the experimental value of $\alpha_3$.
In fact $m_O$ is predicted, once $m_T$ is fixed by the requirement of reproducing the correct DM relic density
(see section \ref{pheno}).
For $m_T \simeq 2.7$ TeV ($100$ GeV) one gets at one loop  $\alpha_{GUT}\simeq 1/39$,
$M_{GUT}\simeq 1.5\cdot 10^{15}$ GeV ($3.1\cdot 10^{15}$ GeV)
and  $m_O \simeq 7\cdot 10^{10}$ GeV ($2\cdot 10^{9}$ GeV).
These heavy colour octets may have some consequences for cosmology, if the reheating temperature is larger than their mass.
Once they are thermally produced, their energy density decreases through various stages of annihilations 
(see the discussion in Ref.~\cite{AD}). The relic colour octets then decay into $T$ plus SM particles 
via higher dimensional operators suppressed by the GUT scale. We checked that, for $m_O$ as large as required by unification,
their abundance and lifetime are small enough to satisfy easily all experimental constraints \cite{AD}.

In section \ref{trip} we will show that  simple $SO(10)$ models may account for this mass hierarchy, which is actually similar
to the one predicted by the class of $SU(5)$ neutrino mass models with a fermionic $24$ multiplet \cite{BS,DF,BNS,FIR}.\footnote{The
mass spectrum of these models contains a few other multiplets below the GUT scale besides $T$ and $O$.
The minimal renormalizable version of these models \cite{pavel} contains even more light multiplets and thus different possibilities to
realize gauge coupling unification.}
It may be worth to compare the present scenario with the case of split supersymmetry \cite{AD}. There unification is achieved adding at 
TeV scale the wino $T$, the gluino $O$, as well as the higgsinos $2_{\pm 1/2}$, and further taking the sfermion mass scale $\tilde{m}$
well below $M_{GUT}$, $\tilde{m}\lesssim 10^{11}$ GeV \cite{GR}.
Other sets of TeV scale fields including $T$ as possible DM candidate and leading to gauge unification are discussed in 
Refs.~\cite{ma,tec}.
To the best of our knowledge, all previous models contain extra multiplets beside $T$ at the TeV scale and, moreover, $T$ is not
automatically stable.

\section{Phenomenology of the fermion triplet \label{pheno}}

Let us discuss in some detail the phenomenology of the fermion isotriplet  $T$ with zero hypercharge as DM candidate.

\underline{Relic density}: In our framework matter parity guarantees that $T$ has no interactions with SM particles beside 
the weak gauge interaction.
It behaves as a wino in the limit where all other superpartners are much heavier. 
Gauge interactions lead to a mass splitting between the charged and neutral component, $m_{T^+}-m_{T_0}=166$~MeV \cite{CDM}, 
independent of the triplet mass, so that the DM is the neutral component as it must.
The triplet mass is fixed by the requirement
to reproduce the observed DM relic abundance and it was accurately computed in Refs.~\cite{Cirelli:2005uq,H06,CST}. The result is 
$m_T = 2.75\pm0.15$ TeV,
where the error accounts for the present $3\sigma$ uncertainty on $\Omega_{DM}$: $0.095< \Omega_{DM} h^2 < 0.125 $. 
For this calculation the mass splitting can be neglected since, at the  freeze-out temperature, the charged component had 
not the time to decay to the neutral component. 
The relevant annihilation cross section is therefore given by the annihilation and co-annihilation of all triplet components,
which gives $\langle \sigma v \rangle=37 g^4/(96 \pi m^2_T)$ \cite{Cirelli:2005uq}. 
The value of $m_T$ given above takes into account the effect of the Sommerfeld enhancement of the cross section stemming from the fact that the triplets are non-relativistic when they freeze-out.
This effect shifts the mass from 2.4~TeV to 2.75~TeV \cite{H06,CST}.
Note that extra interactions of $T$ with other new TeV scale particles would increase the annihilation cross section 
$\langle \sigma v \rangle$ and thus need to be compensated by a larger mass, since 
$\langle \sigma v \rangle \propto 1/m_T^2$.

Smaller values of $m_T$ are possible if the fermion triplets account only for a fraction of the DM energy density.
A non thermal scenario in some cases may also allow a smaller DM mass
but at the price of loosing predictivity.
In the following we do not consider these scenarios which could reduce the mass
and keep $m_T=2.75$~TeV. Still it is interesting to keep in mind that possibilities of this kind do exist.

\underline{Proton decay}:
As discussed in section \ref{unif}, once $m_T=2.75$ TeV is fixed by the DM amount, 
the one-loop estimate for the scale where $\alpha_1$ and $\alpha_2$ meet is $M_{GUT}\simeq 1.5\cdot 10^{15}$ GeV,
where we assume that no extra multiplet with $b_1^{(a)}-b_2^{(a)}\ne0$ is present below the GUT scale, except for $T$.
If one ignores for the moment GUT thresholds, $M_{GUT}$ can be identified with the mass of the
GUT gauge bosons which mediate proton decay.

The most stringent bound comes from the decay $p\rightarrow e^+\pi^0$, whose lifetime can be written as 
\begin{eqnarray}
  \tau (p \rightarrow \pi^0 e^+)\ \simeq\ \left(8.2 \times 10^{33}\, \mbox{yrs} \right)
    \left( \frac{2.3}{A_{SD}} \right)^{\! 2}
    \left( \frac{1/39}{\alpha_{\rm GUT}} \right)^{\! 2}
    \left( \frac{M_V}{4.3\cdot 10^{15}\, \mbox{GeV}} \right)^{\! 4} ,  \hskip .5cm
\label{D=6}
\end{eqnarray}
where we factored out the present $90\%$ C.L. lower bound by Super-Kamiokande \cite{SK}.
For simplicity we included only  the contribution of the $SU(5)$ gauge bosons $(3,2,-5/6)$, with mass $M_V$, which is
a good approximation if the $(3,2,1/6)$ gauge boson mass $M_{V'}$ 
is slightly larger (the lifetime decreases roughly by a factor $4/5$ for $M_{V'}=2M_V$).
The parameter $A_{SD}$ accounts for the renormalization of the
four-fermion operator $(u_R d_R)(u_Le_L)$ from $M_{GUT}$ to $m_{Z}$ and is given by \cite{BEGN}
\begin{equation}
  A_{SD}\ \simeq\ 
    \left( \frac{\alpha_1 (m_Z)}{\alpha_{GUT}} \right)^{- \frac{11}{20 b_1}}
    \left( \frac{\alpha_2 (m_Z)}{\alpha_{GUT}} \right)^{- \frac{9}{4 b_2}}
    \left( \frac{\alpha_3 (m_Z)}{\alpha_{GUT}} \right)^{- \frac{2}{b_3}}\, ,
\label{asd}\end{equation}
where Yukawa contributions to the running have been neglected. In deriving Eq.~(\ref{D=6}) we took the same
renormalization factor also for the other four-fermion operator, $(u_Ld_L)(u_Re_R)$ (they differ only for
the $U(1)_Y$ exponent, $-23/20$ instead of $-11/20$ \cite{BEGN}). Here $b_i$ are the total
$\beta$-function coefficients at $m_Z$, that we assume to include the SM plus $T$,
and thus we obtain $A_{SD}\simeq 2.3$.\footnote{In principle the $T$ and later the $O$ 
contributions to the $b_i$ coefficients should enter in Eq.~\ref{asd} at the scale of their mass; we
checked that this amounts to increase $A_{SD}$ by less than 5\%.}

This analysis shows that the one-loop prediction for the GUT scale, $M_{GUT}\simeq 1.5\cdot 10^{15}$ GeV,
is about three times smaller than the lower bound on gauge boson masses imposed by proton decay,
$M_V\gtrsim 4.3 \cdot 10^{15}$ GeV. The two-loop analysis has been performed for a similar mass spectrum,
in the $SU(5)$ model with a fermionic $24$ multiplet \cite{BNS}. In this case the value of $M_{GUT}$ can
be a factor of 2 larger than the one-loop value, and we expect an analog effect in the present case, 
i.e. $M_{GUT}\sim 3\cdot 10^{15}$~GeV.
Another factor that may affect $M_{GUT}$ by a factor of a few are GUT thresholds, that are expected to be
sizable in realistic models of $SO(10)$ symmetry breaking. One possibility is that extra states with $b^{(a)}_1-b^{(a)}_2 \ne 0$
are lighter than $M_{GUT}$, for example the $(3,2,1/6)+(\bar{3},2,-1/6)$ fermion multiplets already discussed in section \ref{unif}.
If their mass is lowered to $10^{14}$ GeV, the GUT scale is raised to $M_{GUT}\simeq 5\cdot 10^{15}$ GeV.
Another possibility is to break $SO(10)$ to the Pati-Salam subgroup at a slightly larger scale, 
thus giving mass to the gauge bosons responsible for $p$-decay,
and to break Pati-Salam to the SM at a smaller scale.\footnote{For a recent analysis
of non-supersymmetric $SO(10)$ models with intermediate symmetry breaking scales see Ref.~\cite{BDM}.
The proton decay rate in this context was discussed e.g. in Ref.~\cite{LMPR}.}
Finally, it should be kept in mind that Eq.~(\ref{D=6}) also assumes minimal $SU(5)$ Yukawa coupling of light fermions,
which is not the case in realistic models of fermion masses. In fact, the freedom in the Yukawa couplings
can be used to suppress drastically some $p$-decay channels, leading to a much weaker bound 
on the GUT gauge boson mass, $M_V\gtrsim 10^{14}$ GeV \cite{DF2}.

All in all, the fermion triplet DM scenario predicts gauge-mediated proton decay close to the present experimental bound.
However a precise estimate of the proton lifetime requires to specify an explicit model for the $SO(10)$ symmetry breaking 
and the fermion mass generation.

\underline{Direct DM searches}: Since the $T^0$ does not couple in pairs, neither to the Higgs nor to the $Z$ boson,
there is no elastic scattering with a nucleon at tree level. At one loop this process can occur through diagrams involving
two virtual $W$'s scattering off a quark, see Fig.~1 of Ref.~\cite{Cirelli:2005uq}, leading to a suppressed 
spin-independent cross section two-three 
orders of magnitudes below the actual experimental sensitivities, but within reach of the planned sensitivity of future
experiments \cite{CDMSetal}. The inelastic scattering with a charged component is kinematically forbidden because
the $T^+ - T^0$ mass splitting is about three orders of magnitudes above the DM kinetic energy and also above the proton-neutron mass difference.

\underline{Indirect DM searches}: ``Today" the DM annihilations at tree level are to a $W^\pm$ pair whereas at one loop
they can also proceed to $\gamma \gamma$, $\gamma Z$ and $ZZ$. The corresponding positron, antiproton and photon fluxes, 
both diffuse and from the center of the milky way have been determined in Refs.~\cite{H04,Cirelli:2005uq,CST,CFS}. 
Given the fact that the $DM$ is highly non relativistic today, and since 
$m_T\simeq 2.7$~TeV is not far from the value $m_\star=2.5$~TeV where a 
Sommerfeld resonance occurs, a significant boost is induced for these annihilations. 
This would go in the right direction (but is not sufficient by itself)
to explain the positron excess observed by the Pamela experiment \cite{pamE}. In any case,
given the energies of the $W$, $E_W\sim m_{DM}$,
a large enough positron flux would unavoidably lead to a large excess of antiprotons with energies below 100 GeV,
where no excess has been observed \cite{pamP}.

It appears more promising to search for the annihilations $TT\rightarrow \gamma\gamma$, leading to monochromatic 
photons with energy $m_T$, which are also Sommerfeld enhanced and well within the reach of atmospheric Cherenkov
telescopes looking at the galactic center \cite{H04,CFS}. The non-observation of this signal may rule out $T$ as DM candidate,
while a positive signal may allow a direct determination of its mass.

Note also that since the matter parity is a subgroup of $SO(10)$, we do not expect that high scale physics could cause any decay of the triplet.

\underline{Collider signatures}: 
The possibility to observe a $Y=0$ fermion isotriplet at the LHC has been studied in details 
in Refs.~\cite{FranceschiniHS,daas}.\footnote{The LHC phenomenology in the analog supersymmetric case, 
with the wino as the lightest superparticle, has been studied e.g. in Ref.~\cite{IMY}.}
It appears to be possible for a mass up to 1.5~TeV.
For a mass of $2.7$~TeV and
an integrated luminosity of $100$~fb$^{-1}$ (which is roughly the one that each detector is  expected to collect in one year with the
full LHC luminosity), the $p p \rightarrow DM\,DM\,X$ production cross section (with $X$ any other particle) leads 
only to about one produced DM pair (see also Refs.~\cite{Cirelli:2005uq,MR}). Possible future upgrades of the LHC luminosity are discussed in 
Ref.~\cite{LHCupgrades}. A hadronic collider with twice more energy would lead to a production cross section several orders of magnitudes larger.
Also
an $e^+ e^-$ collider with an energy above twice the DM mass would allow its observation, producing a $T^+ T^-$ pair through a $Z$ at tree level, or a $T^0 T^{0*}$ DM pair through a one loop box diagram with 2 virtual $W$'s.

If produced, the triplet displays a clean signature in the form of long lived charged tracks
as the lifetime of the charged components (from $T^\pm \rightarrow T^0 \pi^\pm,\, T^0 {l^\pm} 
^{\scriptscriptstyle(}\bar\nu_l ^{\scriptscriptstyle)}$ decays) 
is definitely predicted by the gauge interactions, $\tau_{T^\pm}\simeq 5.5$~cm \cite{Cirelli:2005uq,FranceschiniHS}.
Contrary to the case of Refs.~\cite{FranceschiniHS,daas} where $T^0$ can decay into leptons, in our scenario it is completely
stable because of matter parity, therefore its production will manifest as missing energy.
Effects of a 2.7~TeV triplet on electroweak precision data are negligible.

\underline{Neutrino masses and baryogenesis}:
The obvious source of Majorana neutrino mas\-ses in our scenario is the type I seesaw, since $SO(10)$ models contain automatically right-handed
neutrinos. Note that they do not affect the gauge unification analysis at one loop.
Moreover, they may account for the baryon asymmetry of the Universe through the leptogenesis mechanism.
The triplet $T$ does not mediate neutrino masses because
the exact matter parity $P_M$ prevents the coupling $y_\nu l\,T\,h$, where $l$ ($h$) is the SM lepton (Higgs) doublet.
Notice that the very small Yukawa coupling $y_\nu$, required to generate neutrino masses in the case of a TeV scale triplet, 
would be nonetheless too large to preserve the DM stability on cosmological time scales.

\section{$SO(10)$ models with a light fermion triplet  \label{trip}}

In unified models the mass of vector-like fermions is not bound to the electroweak scale,
contrary to the case of the chiral SM fermions. The fermion DM candidates under consideration
are vector-like and a special mechanism seems to be required
to lower their mass $m_{DM}$ much below the GUT scale, a problem analog to
the well-known doublet-triplet splitting problem of supersymmetric unified models.
Notice that the smallness of $m_{DM}$ is technically natural in the effective theory below $M_{GUT}$, because when it tends to zero 
one recovers an extra $U(1)$ global symmetry.\footnote{An analogous situation has been analyzed for  
models with supersymmetry broken at large scale but light gauginos and higgsinos (split supersymmetry \cite{AD,GR}).
An additional $U(1)$ symmetry is recovered in the limit where both gauginos and higgsinos are massless.} 
Then, it may also be natural in the full theory, if the GUT scale physics respects such a global symmetry.
In this section
we investigate possible mechanisms to lower the triplet mass $m_T$ to the TeV scale in $SO(10)$ models. 
In view of the unification constraint, we will also demand a colour octet fermion $O$ at intermediate scale.

Let us discuss first the possibility to achieve the smallness of $m_T$ and $m_O$ by a 
fine-tuning of the $SO(10)$ couplings, while all the other components of the same $SO(10)$ multiplet receive a
GUT scale mass.
We assume for definiteness that $T$ and $O$ belong to a $45$ fermion multiplet.
The simplest  way to lower their masses is to introduce three (or more) couplings contributing  with
different Clebsch-Gordan coefficients to the masses of the various $45$ components, for example
\begin{equation}
45(M+y_{54} 54_H+y_{210} 210_H)45 ~,
\label{ren}\end{equation}
where the SM singlets in the Higgs multiplets $54_H$ and $210_H$ have a VEV at the GUT scale. In this case $m_T$ and $m_O$ are given by 
two linear combinations of $M$ and the VEVs that can be both tuned to the small values required by dark matter and unification constraints, 
while all the  other components of $45$ live close to $M_{GUT}$.

To avoid large Higgs representations like $210_H$, one may include instead dimension five operators. Sticking to the $54_H$
multiplet, one has
\begin{equation}
\frac{1}{\Lambda}\left[
c_1 Tr(45\,45)Tr(54_H\,54_H) + c_2 Tr(45\,54_H\,45\,54_H) + c_3 Tr (45\,45\,54_H\,54_H) 
\right]~. 
\label{nonren}\end{equation}
This is the $SO(10)$ embedding of a $SU(5)$ neutrino mass model presented in Ref.~\cite{BS}, 
with a fermionic adjoint $24^{SU(5)} \subset 45$ and a Higgs adjoint $24_H^{SU(5)} \subset 54_H$.
It was shown that this setting is sufficient to lower
the masses of $T$ and $O$ contained in the $24^{SU(5)}$ fermion multiplet, at the price of fine-tuned cancellations 
between the renormalizable 
and non-renormalizable terms, that is to say, in the present $SO(10)$ embedding,
between the $M$ and $y_{54}$ terms in Eq.~(\ref{ren}) and those in Eq.~(\ref{nonren}).

A more ambitious goal is to forbid the GUT scale masses of $T$ and $O$ by a symmetry and
thus recover the desired mass spectrum without fine-tuning.
It may be worth to make a comparison with
the doublet-triplet splitting problem in supersymmetric GUT models, where a light pair of Higgs chiral superfields can
be obtained either by fine-tuning or by a dynamical mechanism, that requires a specific structure 
of the superpotential, which may be justified by some global symmetry.
There is of course one important difference: 
no matter how the small mass is realized, supersymmetry guarantees that its smallness is radiatively stable,
while in the non-supersymmetric case a global symmetry is a requisite to maintain naturalness.

To justify the lightness of $T$ with a symmetry, we develop a model based on a variation of  
the missing VEV mechanism for doublet-triplet splitting \cite{DW}.\footnote{The 
original mechanism can be applied straightforwardly to our non-supersymmetric framework
in the case of the fermion DM candidates $2_{\pm 1/2}$, belonging to a $10$ multiplet.
It is enough to forbid all mass terms for the $10$ except the coupling
$10~ 10'~ 45_H^{B-L}$, where $45_H^{B-L}$ is an adjoint Higgs with VEV in the $B-L$ direction.
Then the isodoublets in $10$ do not acquire any GUT scale mass.}
For this purpose consider the lagrangian 
\begin{equation}
-{\cal L} = y_{12}\,45_1\,45_2\,45_H^Y + \frac 12 M_2\,45_2\,45_2 + h.c. ~.
\label{reno}\end{equation}
The $45_H^Y$ is assumed to acquire a VEV in the hypercharge direction $T_Y$, that is,
the SM singlet $S^Y_H \subset 24^{SU(5)}_H\subset 45_H$ has a non-zero VEV.
In this case the Yukawa coupling $y_{12}$ provides a mass to all the fermions in $45_1$ and $45_2$, 
except the isotriplets $T_{1,2}$, the colour octets $O_{1,2}$ as well as the singlets $S_{1,2}^X$ and $S_{1,2}^Y$, 
contained in the $SU(5)$ singlet and adjoint component of $45_{1,2}$, respectively.\footnote{To check this, recall 
that the $SO(10)$ contraction of three adjoints is completely antisymmetric.
If only $S^X_H \subset 1^{SU(5)}_H\subset 45_H$ acquired a VEV, additional components of $45_{1,2}$ would remain massless.
The same problem occurs if the VEV is in the $T_{3R}$ or $T_{B-L}$ direction only, that are special linear combinations of $T_X$ and $T_Y$.}
The mass term $M_2$ makes all $45_2$ components heavy.
This is a promising first step toward the desired mass spectrum and can be easily justified by a global symmetry,
with $45_1$ and $45_H^Y$ carrying opposite charges and no charge for $45_2$.
It is remarkable that exactly the multiplets required for dark matter and gauge unification, $T_1$ and $O_1$, 
do not receive a mass at the GUT scale.

The second step is to generate the appropriate contributions to the masses of $T_1$ and $O_1$.
Since we need $m_{O} \sim 10^{10}$ GeV and $m_{T} \sim 10^3$ GeV $\sim m_{O}^2 / M_{GUT}$,
an interesting possibility is to introduce a unique intermediate mass scale $m_{int}$ and suppress $m_{T}$ by a seesaw-like mechanism.
The required mass textures  for the triplet and octet components in the basis $(45_1,45_2)$ are the following:
\begin{equation}
{\cal M}_T \sim \left(\begin{array}{cc}
0 & m_{int} \\ m_{int} & M_2
\end{array}\right) ~,~~~~~
{\cal M}_O \sim \left(\begin{array}{cc}
m_{int} & (m_{int}) \\ (m_{int}) & M_2
\end{array}\right) ~.
\label{seeT}\end{equation}
where $M_2\sim M_{GUT}$, $m_{int}\sim 10^{10}$ GeV and unnecessary entries are put in brackets. 
To achieve this pattern requires some model-building. The simplest way to provide different contributions to triplet and octet masses
is the coupling $45\; 54\; 45_H$: when the $45_H$ VEV lies in the right-handed isospin direction $T_{3R}$, the fermion colour
octets in $45$ and $54$ do not acquire a mass, while the isotriplets do; the opposite happens when the VEV direction is $T_{B-L}$.
This indicates the need to introduce the additional Higgs multiplets $45_H^{B-L}$ and $45_H^{3R}$,
where the superscript specifies the VEV alignment, as well as to impose appropriate global symmetries to forbid unwanted mass terms.
The field content of a minimal model and the global charge assignments are shown in Table \ref{model}.

\begin{table}[t]
\begin{center}
\begin{tabular}{|c|cc|ccc|}
\hline
& \multicolumn{2}{|c|}{fermions} & \multicolumn{3}{|c|}{scalars} \\ 
& $45_1$ & $45_2$ & $45_H^Y$ & $45_H^{3R}$ & $45_H^{B-L}$ \\
\hline
$Z_2$ & $-$ & $-$ & $+$ & $+$ & $-$ \\
\hline
$Z_4$ & $+i$ & $-1$ & $+i$ & $+1$ & $-i$ \\
\hline
\end{tabular}
\end{center}
\caption{\it{\small{The fermion and Higgs multiplets in the model of Eqs.~(\ref{reno}) and (\ref{nonreno}) and their charges
under a global symmetry $Z_2\times Z_4$, chosen to realize the desired mass spectrum.}}
\label{model}}
\end{table}

The lagrangian invariant under $SO(10)$ and the global symmetries includes the renormalizable
terms in Eq.~(\ref{reno}) plus the following dimension five operators:
\begin{equation}
-{\cal L}^{non-ren} = \frac{c_{11}}{\Lambda} (45_1 45_H^{B-L})_{54} (45_H^{B-L}45_1)_{54}
+ \frac{c_{12}}{\Lambda} (45_1 45_H^{3R})_{54} (45_H^Y 45_2)_{54} + h.c. ~.
\label{nonreno}\end{equation}
The subscript $54$ specifies the contraction of the $SO(10)$ indexes of each pair of $45$'s:
we assume that 
$54$ fermion multiplets with the appropriate global charges live at the cutoff scale $\Lambda$. 
The operator $c_{11}$ fills the $11$ entry in the colour octet mass matrix ${\cal M}_O$ in Eq.~(\ref{seeT}). 
The operator $c_{12}$ contributes to the 
off-diagonal entries in ${\cal M}_T$.
The split of octet and triplet masses is thus generated by higher dimensional operators, that can be suppressed by a $\Lambda$ as large
as the Planck scale, $M_P \sim 10^3  M_{GUT}$.
This partially accounts for the smallness of $m_{int}$ with respect to $M_{GUT}$.
To obtain the required values of $m_T$ and $m_O$ one further needs $c_{11}\sim 10^{-2}$ and  $c_{12}\sim 10^{-3}$.

This schematic model may not be the simplest possibility, 
yet it proves that the ordinary  $SO(10)$ properties
allow to split the isotriplet and colour octet masses from the other components
of the GUT multiplets. 
The large hierarchy among these masses may indicate that an additional global 
symmetry suppresses the values of $c_{11}$ and $c_{12}$. 
We remark that the need of some (approximate) global symmetries, like those in Table \ref{model},
does not change the status of the matter parity $P_M$
as an exactly conserved subgroup of the unified gauge symmetry. 

We note, in addition, that light fermion multiplets can be simply obtained
in a less conventional class of GUT models, in which the GUT symmetry is broken by orbifold
compactification \cite{orbi}.
By an appropriate choice of boundary conditions, only some fragments of the unified multiplets possess a zero mode.
This property has been used in the supersymmetric case to solve the doublet-triplet splitting problem \cite{K2}.

\section{Conclusions}

We considered the possibility to realize grand unification in the SM augmented by a TeV scale candidate
for the DM. The stability of the DM can be an automatic consequence of the unified gauge symmetry;
this is actually a powerful criterion  to select a DM candidate.
Another criterion is to realize gauge coupling unification above the lower bound imposed by the proton
decay. We found 
that both criteria, as well as constraints from direct and indirect searches, can be satisfied at the same time.
The simplest candidate for this unified dark matter (UDM) scenario is
a fermion isotriplet with no hypercharge, $T$. 
We sketched a few non-supersymmetric $SO(10)$ models in which $T$ can be made much lighter than the GUT scale.
Barring extra contributions to $\Omega_{DM}$ or non-thermal scenarios, the relic density constraint fixes its mass to about $2.7$ TeV.
The UDM is thus slightly beyond the LHC reach, but can be observed by future direct DM searches,
as well as through its annihilations into monochromatic gamma rays.
The predicted value of $M_{GUT}$ is close to the lower bound imposed by $\tau(p\rightarrow e^+\pi^0)$.

\section*{Acknowledgements}

MF thanks the Service de Physique Th\'eorique de l'Universit\'e Libre de Bruxelles
for the kind hospitality  during the accomplishment of this work. The work of MF was supported in part
by the Marie-Curie Intra-European Fellowship MEIF-CT-2007-039968, the CICYT Research Project
FPA-2008-01430, and the Marie-Curie Research and Training Network ``UniverseNet'' MRTN-CT-2006-035863.
The work of TH is supported by 
the FNRS-FRS, the IISN and the Belgian Science Policy (IAP VI-11).



\end{document}